\documentclass[10pt, a4paper, onecolumn]{article}

\usepackage[numbers]{natbib}

\usepackage[T1]{fontenc}
\usepackage{lmodern}

\usepackage{natbib}
\usepackage[utf8]{inputenc} 
\usepackage{booktabs}       
\usepackage{amsfonts}       
\usepackage{nicefrac}       
\usepackage{microtype}      
\usepackage{xcolor}         
\usepackage{graphicx}
\usepackage{amsmath,amssymb,amsfonts}
\usepackage{bm}
\usepackage{url}

\usepackage{array,multirow,graphicx}

\usepackage{algorithm}
\usepackage[noend]{algpseudocode}

\usepackage{geometry}
\newgeometry{vmargin={15mm}, hmargin={12mm,17mm}}

\title{HPCGen: \underline{H}ierarchical K-Means Clustering and Level Based \underline{P}rincipal \underline{C}omponents for Scan Path \underline{Gen}aration}

\author{
	Wolfgang Fuhl\\
	Department of Human Computer Interaction\\
	University Tübingen\\
	Tübingen, 72076 \\
	\texttt{wolfgang.fuhl@uni-tuebingen.de} \\
}

\begin{document}

	\maketitle
	
	\begin{abstract}
		In this paper, we present a new approach for decomposing scan paths and its utility for generating new scan paths. For this purpose, we use the K-Means clustering procedure to the raw gaze data and subsequently iteratively to find more clusters in the found clusters. The found clusters are grouped for each level in the hierarchy, and the most important principal components are computed from the data contained in them. Using this tree hierarchy and the principal components, new scan paths can be generated that match the human behavior of the original data. We show that this generated data is very useful for generating new data for scan path classification but can also be used to generate fake scan paths.\\
		\url{https://atreus.informatik.uni-tuebingen.de/seafile/d/8e2ab8c3fdd444e1a135/?p=%2FHPCGen&mode=list}.
	\end{abstract}

	\section{Introduction}
	
	\begin{figure}
		\centering
		\includegraphics[width=0.6\textwidth]{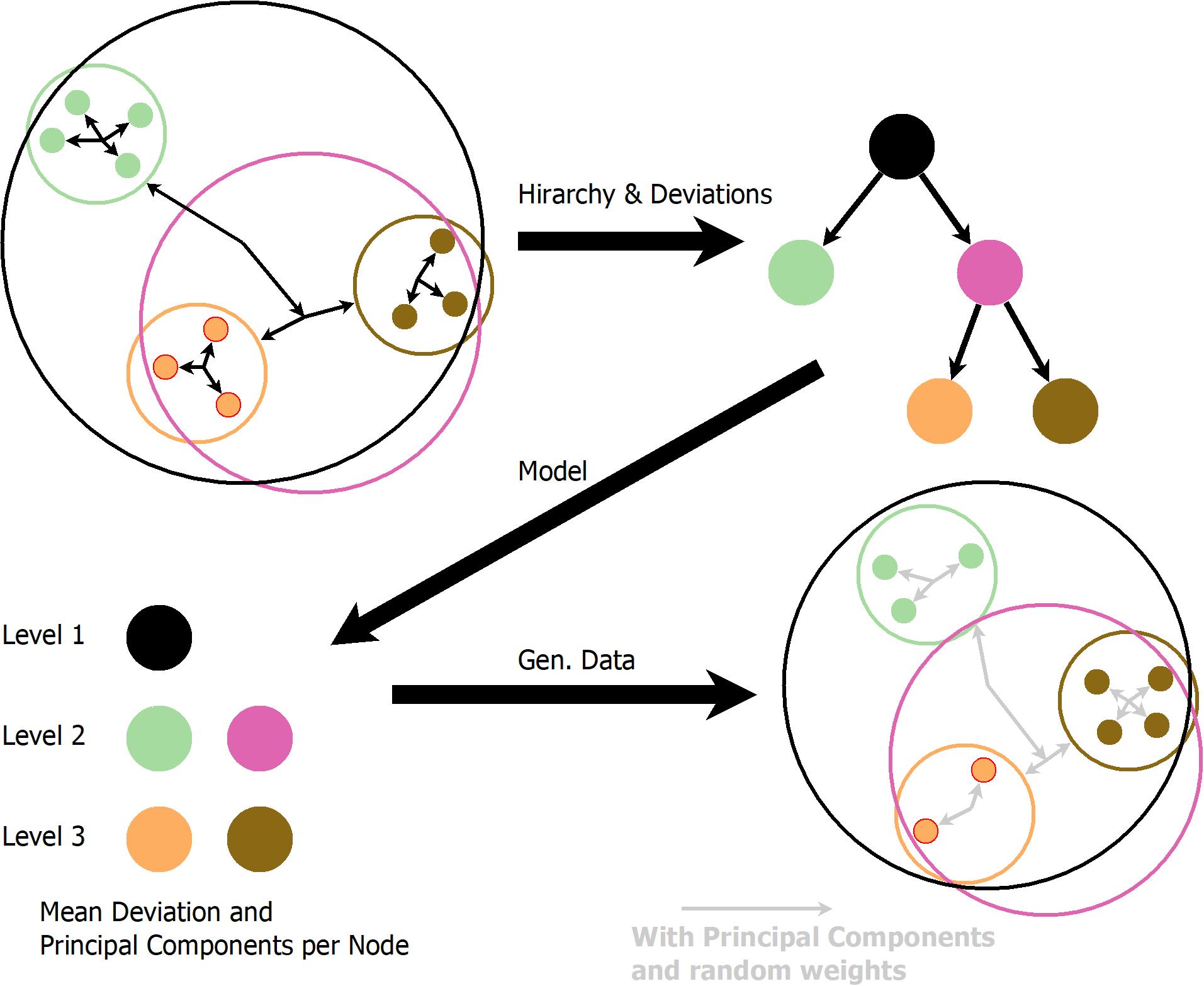}
		\caption{The general concept of our proposed approach. We use multiple gaze recordings and cluster them hierarchically. In each cluster, we compute the shift vectors to the cluster center of the data points and compute the next level with the clustering of the deviations. Based on the shifts we compute the principal components which we can afterwards use to generate a scan path. The shown example uses only two clusters per level, our approach can use different amounts of clusters on each level.}
		\label{fig:teaser}
	\end{figure}

	In today's world, eye tracking can be found in many fields. This includes human-machine interaction~\cite{gardony2020eye,arslan2021eye}, computer graphics~\cite{walton2021beyond,meng2020eye}, self-diagnosis systems ~\cite{joseph2020potential,snell2020assessment,lev2020eye}, measurement of eye misalignment~\cite{nesaratnam2017stepping,economides2007ocular}, detection of neurological diseases~\cite{davis2020eye,pavisic2021eye}, behavioral research~\cite{lewandowski2020factors,chauliac2020all}, learning systems~\cite{panchuk2015eye,jermann2010using}, expertise determination~\cite{brunye2020eye,richter2020massed}, driver observation~\cite{liu2019gaze,shinohara2017visual,NNETRA2020}, market research~\cite{bialkova2020desktop,sielicka2021consumer}, and many more.
	
	In computer graphics, for example, the main interest is to render only the 2 degrees of the fovea with a high resolution~\cite{walton2021beyond,meng2020eye}. In the rest of the image it is sufficient to render roughly, which would result in a significant reduction of computational effort~\cite{walton2021beyond,meng2020eye,FFAO2019}. One problem that exists here is the need for many scan paths, which is necessary for testing the system. In the area of market research, one evaluates product placements or the online presence of manufacturers and sellers~\cite{bialkova2020desktop,sielicka2021consumer}. Here, many scan paths are also needed to make a general statement or a statement about specific customer groups. This is also neccesary for an analyisis of the eye tracking data~\cite{AGAS2018,ROIGA2018,ASAOIB2015}. In self-diagnostic systems, a disease or defect is diagnosed based on the user's behavior~\cite{joseph2020potential,snell2020assessment,lev2020eye,davis2020eye,pavisic2021eye,UMUAI2020FUHL}, which can then later be treated or examined by a doctor. For these systems to work reliably, it is also necessary to use a lot of data and many different features like the eye lid closere~\cite{WTDTWE092016,WTDTE022017,WTE032017}, pupil variations~\cite{WTCKWE092015,WTTE032016,062016,CORR2017FuhlW2,CORR2016FuhlW,WDTE092016} or other features extracted from the eye~\cite{WTCDAHKSE122016,WTCDOWE052017,WDTTWE062018,VECETRA2020,ETRA2018FuhlW,ETRA2021PUPILNN,ICCVW2019FuhlW,CAIP2019FuhlW,ICCVW2018FuhlW}. In the area of driver observation, the same applies. Here it is necessary to detect inattentive behavior or drowsiness and warn the driver~\cite{liu2019gaze,shinohara2017visual}. These systems also need to be tested on a lot of data to ensure their reliability and optimize their usability for the driver. In the field of human-machine interaction, many data are also needed, which are used to test interactions and make the final algorithm as reliable as possible~\cite{gardony2020eye,arslan2021eye,ICML2021DSISMAR,CORR2017FuhlW1}. Thus, it is clear that data is becoming increasingly important in the machine learning era. Since the recording and annotation of data is very time-consuming, and this data is subject to data protection, science has long been concerned with the simulation of data and its use for the validation~\cite{NNVALID2020FUHL,ICMV2019FuhlW} of systems or for training machine learning methods~\cite{AAAIFuhlW,NNPOOL2020FUHL,NORM2020FUHLICANN,RINGRAD2020FUHL,NIPS2021MAXPROP,2021TNandFDT}.
	
	In the literature, there are of course a variety of algorithms that have dealt with the simulation and generation of data in different data domains~\cite{vilardell2020missing,rasheed2020digital}. However, simulating human behavior remains a challenge to this day. The latest methods use Biological Reference Values, which are used to generate scan paths~\cite{EPIC2018FuhlW}. However, the focus of this work is to generate eye movement types~\cite{MEMD2021FUHL,WF042019} as realistically as possible based on the velocity characteristics. Another approach deals with the generation of scan paths via a variational autoencoder, which learns distribution parameters~\cite{FCDGR2020FUHL}. New data can then be generated using these distribution parameters, but these were only used as an add-on for data augmentation. To hide specific data in a scan path, such as the Personal Features, a reinforcement learning based approach was presented~\cite{RLDIFFPRIV2020FUHLICANN}. Here, the data in the middle part of the autoencoder is manipulated to manipulate the scan path. The focus of this work was to create scan paths that contain only selected information. Another work uses autoencoders together with data augmentation in the target layer of the autoencoder to obtain a representation of a scan path that can be classified with linear models~\cite{bautista2021gazemae}. All the work mentioned here generates scan paths, but never with the focus to simulate realistic human behavior.
	
	In this work, we present an algorithm that allows to learn a scan path model hierarchically from given human scan path images. This means that based on the given scan paths, new scan paths can be generated that reflect exactly this behavior. An example would be the search for items in an online store. If some data from humans is given here, a model can be generated with our algorithm and with this model new scan paths for the search of articles. This can be used to generate data for scan path classification as well as data to validate an online presence. Of course, there are limitations to the application of our algorithm, but these are described in the limitations' section of this paper.

	Our contributions to the state of the art are:
	\begin{enumerate}
		\item A novel approach for scan path generation which is totally data driven.
		\item Code for the generation of scan path, as well as code to train new models.
		\item Comparison to other state of the art approaches, as well as the combination with other approaches to generate training data.
		\item Evaluation of fake scan path generation
	\end{enumerate}
	
	\section{Related work}
	The first part of the related work focuses on the simulation of eye movements. In 2002 an approach on image rendering was proposed in \cite{lee2002eyes}. The authors focused on the generation of realistic saccade movements, and it was also possible to simulate smooth pursuits as well as different vergence angels. Another approach which focuses on the generation of eye movements as well as head rotations was proposed in \cite{ma2009natural}. Here the head movements are simulated additionally, but the simulator automatically triggers an eye movement, if the head motion starts. The next improvement was proposed in \cite{le2012live}. Here, head movements do not trigger eye movements anymore and in addition to the standard eye movement types, blinks are simulated as well. For the generation of the eye movements, the method in \cite{le2012live} uses sound files. There exists also an approach which is motivated by the eye muscle~\cite{tweed1990computing} movement. The approach is described in \cite{duchowski2015modeling} and focuses especially on the eye rotation. This approach uses predefined sequences of eye movements to generate the eye rotations, for which it is not a real eye movement nor a gaze data generation approach. The last approach which is based on image rendering is \cite{wood2015rendering}. This approach randomly generates eye images with different textures as well as different orientations. The main purpose of this approach is to generate training data for computer vision algorithms, since no eye movement types nor realistic gaze data is generated. The so far mentioned approaches are usually used to generate images or sequences of images corresponding to special eye movement types. Their main use is in animation of characters or the interaction with humans~\cite{pejsa2013stylized}. Therefore, the generated eye movement sequences do not correspond to real viewing behavior of humans, since no saliency information nor task specific knowledge is integrated.

	An approach which focuses on a realistic simulation of gaze sequences based on saliency information is described in \cite{campbell2014saliency}. The approach is based on a statistical model and uses saliency information together with random number generators to simulate gaze data. In \cite{duchowski2015eye} the before mentioned approach was refined by adding different distributions for noise generation, which is a part in any real world data recording of gaze data. Additionally, the authors in \cite{duchowski2016eye} added jitter of the gaze signal itself based on a normal distribution. An extension of those approaches which supports different sampling rates as well as dynamic sampling rates, which is usually the case if the cameras have no fixed exposure time, is proposed in \cite{EPIC2018FuhlW}. The approach generates random sequences of eye movement types together with the gaze information and is able to map them to static and dynamic stimuli as well as to other gaze data. The approach itself uses different distributions as well as scientifically based equations for the calculation of the gaze signal and eye movements.

	With the upcoming of deep neural networks and the recent improvements in machine learning, other novel approaches have been proposed recently. A deep recurrent neural network to generate gaze heatmaps was proposed in \cite{simon2016automatic}. It is applied on static images and tries to generate a realistic gaze heatmap. While this was one of the first approaches, it has the limitation, that it only works on already seen images during training. An approach using generative adversarial neural networks (GAN) was proposed in \cite{assens2018pathgan}. Here, the input consists of static images and saliency maps out of which a scan path is generated. An approache for scan path manipulation using an autoencoder and reinforcement learning is described in \cite{RLDIFFPRIV2020FUHLICANN}. The authors did not try to compute a realistic scan path, but to manipulate the scan path in a way that some information is removed while other information is still contained in the scan path. An approach using a variational autoencoder is proposed in \cite{FCDGR2020FUHL}. Here, the autoencoder learns parameters of distributions in the central block of the autoencoder. With a random sequence of numbers and those parameters, a new scan path can be generated. The authors used this simulation technique only for data augmentation, which helped to improve the classification accuracy. In \cite{bautista2021gazemae} an autoencoder which combines different feature extractors (Convolutions with different sizes) was proposed. The authors did not try to generate a realistic sequence but use the central part of the autoencoder as general metric for scan path classification.
	
	\section{Method}
	\begin{algorithm}
		\caption{Model Generation}
		\label{alg:Model}
		\begin{algorithmic}[1]
			
			\Procedure{GenerateModel}{$GazeData,MaxLevel,NumClusters,DynCluster$}     
			\State $Model=0$ 
			\State $Clusters[1]=GazeData$ \Comment{Multiple recordings with two or three dimensions (X,Y,time)}
			\For{$l=1~to~MaxLevel$}
			\If{$DynCluster = True$}
			\State $NumClusters=UpdateRule(NumClusters,l)$
			\EndIf
			\For{$c=1~to~Size(Clusters)$}
			\State $Model[l][c].m=Mean(Clusters[c])$ \Comment{First level is position other levels are deviations}
			\State $SubClusters[c] = KMeans(Clusters[c],NumClusters)$ \Comment{SubClusters has multiple dimensions}
			\State \Comment{The clustering will assigne each gaze point a cluster if there are to few gaze points.}
			\State \Comment{The shifts to those gaze points will be computed and used in the PCA.}
			\State $CombineCloseSubClusters(SubClusters)$  \Comment{Assign clusters between multiple recordings}
			\State $Model[l][c].p=PCA(Mean(SubClusters[c])-Model[l][c].m)$
			\EndFor
			\State $Clusters=ToOneDimension(SubClusters)$
			\EndFor		
			\State \textbf{return}~$Model$
			\EndProcedure
			
		\end{algorithmic}
	\end{algorithm}
	The general idea of our approach is to extract the structure of gaze data and reuse this structure to generate new gaze data (Figure~\ref{fig:teaser}). Gaze data generally consists of eye movements such as fixations and saccades~\cite{duchowski2017eye}. These fixations and saccades form clusters. If one now observes a longer time interval, several fixations form new clusters, which are interesting or important image areas or objects in fixed stimuli or in dynamic scenes. If we follow this formation of clusters over the whole image, a hierarchy of clusters is formed, which becomes deeper and finer. In our case, we go the opposite way and try to find the rough clusters from the whole recordings in the first step and then calculate finer and finer divisions for each cluster. 
	
	If we now look at several images of people, we notice that the same areas are often looked at. These areas are clusters in our hierarchy. Thus, our idea is to assign the clusters according to the hierarchy and position (the position can be two or three-dimensional. Thus, include spatial position and time) between the images to each other. Since each cluster has further sub clusters (In the last level of the hierarchy these are gaze points), we can calculate the displacement of these sub clusters to the parent cluster. Together with the assignment of the clusters, we obtain numerous possible shifts. We consider these displacements as shapes, which can be approximated via an active shape model~\cite{cootes1992active}. This active shape model uses the most important principal components, which can then be used to compute new sub clusters. Once we have calculated the general hierarchy over all images, we can use this hierarchy to generate new gaze data. To do this, we run through the hierarchy from above and select random clusters in each layer, which are positioned based on their parent cluster or the principal components they contain. The leaves of the randomly generated tree are then the gaze points that are generated.

	\begin{algorithm}
		\caption{Scan path Generation}
		\label{alg:Generate}
		\begin{algorithmic}[1]
			
			\Procedure{GenerateScanPath}{$Model,MaxClusters,MaxSubClusters,DynCluster$}     
			\State $GazeData=0$     
			\State $ParentClusterMeans[1]=0$  
			\For{$l=1~to~MaxLevel$}
			\If{$DynCluster = True$}
			\State $MaxClusters=UpdateRule(MaxClusters,l)$
			\State $MaxSubClusters=UpdateRule(MaxSubClusters,l)$
			\EndIf
			\For{$lc=1~to~ParentClusterMeans$}
			\State $LastMean=ParentClusterMeans[lc]$
			\State $NumClusters=Rand(0,MaxClusters)$
			\If{$NumClusters = 0$}
			\State $NewParentClusterMeans[lc][1] = LastMean$ \Comment{Clusters can end before the max level is reached}
			\State \Comment{They won't be touched in the next iteration}
			\Else
			\For{$c=1~to~NumClusters$}
			\State $IDX=Rand(1,size(Model[l]))$
			\State $m=Model[l][IDX].m$ \Comment{Mean shift of the model, first level it is the position.}
			\State $p=Model[l][IDX].p$ \Comment{Principal components to compute new shifts.}
			\State $NumSubClusters=Rand(1,MaxSubClusters)$
			\State $RandWeights=RandMatrix(NumSubClusters,ColumnSize(p))$
			\State $NewParentClusterMeans[lc][c] = LastMean + m + RandWeights*p$ 
			\EndFor
			\EndIf
			\EndFor	
			\State $ParentClusterMeans=ToOneDimension(NewParentClusterMeans)$
			\If{$l = MaxLevel$}
			\State $GazeData=ParentClusterMeans$
			\If{$TimeDependent = True$} \Comment{Estimated automatically on the dimensionality of the means}
			\State $GazeData=NormalizeTime(GazeData)$ \Comment{Necessary if the time was used for clustering}
			\Else
			\State $GazeData=AddTime(GazeData)$ \Comment{Necessary if there is no time given}
			\EndIf
			\EndIf
			\EndFor		
			\State \textbf{return}~$GazeData$
			\EndProcedure
			
		\end{algorithmic}
	\end{algorithm}
	
	In Algorithm~\ref{alg:Model} the conceptual procedure for generating a model according to our approach is described.  As input, it needs the gaze data of several images (GazeData), a maximum depth (MaxLevel), from which the calculation is stopped, an initial number of clusters (NumClusters), and whether the number of clusters should change over the depth or not (DynClusters). The gaze data can be passed both as simple x,y coordinates or additionally with the time, which is then considered with the KMeans procedure for the computation of the clusters. In the top level, an initial mean position is determined, from which the displacements for further clusters are then calculated. As soon as the clusters are calculated on all data, these are assigned between the images in order to be able to calculate general information for possible displacements via the sub clusters. In the following step, these possible shifts are divided into the most important subcomponents using the principal component analysis. These principal components are stored in the model together with the mean displacement (in the first layer the mean position). For each further layer in the model, this procedure is repeated until the last layer is reached or there is too little data in each cluster to form new clusters.
	
	Our procedure to use a model to generate new gaze data, is described in Algorithm~\ref{alg:Generate}. As input, this algorithm has, the Model (Model), the maximum number of clusters (MaxClusters), the maximum number of subclusters per cluster (MaxSubCluster), and whether the maximum values for the clusters should change based on the current level (DynCluster). Our algorithm starts at the first level and randomly determines how many clusters to create. For each cluster to be created in this level, the number of subclusters is determined randomly. The displacement of these subclusters is calculated using the principal components and a random matrix. Together with the mean displacement or mean position for the first layer from the previous cluster, the positioning of the subclusters is determined. Then, the current clusters are replaced with the new clusters and for each new subcluster all the above operations are performed again. In the end, the leaves of this tree are obtained, which are the created gaze points. In case no time has been used in the data, the time is added linearly from the beginning to the end. If time has been given in the data, the data is sorted based on the created gaze points and normalized from zero to one.

	\section{Evaluation}

	\def\arraystretch{0.8}
	\begin{table}[htb]
		\setlength\tabcolsep{1pt}
		\caption{The results of our approach in combination with others to improve the classification accuracy of participants and stimuli on two public data sets. Abbreviations: NN is neural network, ET is ensemble of bagged trees.}
		\label{tbl:evalCh}
		\centering
		\begin{tabular}{lcccccccc}
			\toprule
			& & & & & \multicolumn{2}{c}{ETRA2019Challenge Data~\cite{otero2008saccades,mccamy2014highly}} & \multicolumn{2}{c}{Hollywood Data~\cite{costela2019free}}\\
			& \multicolumn{4}{c}{Training Data} & \multicolumn{2}{c}{Accuracy} & \multicolumn{2}{c}{Accuracy}\\
			Method & Sim~\cite{fuhl2018simarxiv} & VAE~\cite{FCDGR2020FUHL} & Proposed & Real Data & Stimulus & Participant & Stimulus & Participant\\
			\midrule
			Encodji~\cite{C2019} & \multirow{3}{*}{X} & & & & 52\% & 23\% & 35\% & 23\% \\
			NN\&Heatmap~\cite{GMS2021FUHL} & & & & & 60\% & 29\% & 31\% & 28\% \\
			ET\&Heatmap~\cite{GMS2021FUHL} & & & & & 61\% & 29\% & 30\% & 31\% \\ \hline
			Encodji~\cite{C2019} &  & \multirow{3}{*}{X} & & & 51\% & 24\% & 36\% & 22\%\\
			NN\&Heatmap~\cite{GMS2021FUHL} & & & & & 61\% & 28\% & 37\% & 34\%\\
			ET\&Heatmap~\cite{GMS2021FUHL} & & & & & 60\% & 31\% & 38\% & 33\% \\ \hline
			Encodji~\cite{C2019} & & & \multirow{3}{*}{X} & & 58\% & 28\% & 39\% & 37\% \\
			NN\&Heatmap~\cite{GMS2021FUHL} & & & & & 64\% & 35\% & 40\% & 35\% \\
			ET\&Heatmap~\cite{GMS2021FUHL} & & & & & 62\% & 37\% & 42\% & 33\% \\\hline
			Encodji~\cite{C2019} & & & & \multirow{3}{*}{X} & 89\% & 81\% & 73\% & 67\% \\
			NN\&Heatmap~\cite{GMS2021FUHL} & & & & & 81\% & 73\% & 59\% & 41\% \\
			ET\&Heatmap~\cite{GMS2021FUHL} & & & & & 82\% & 77\% & 62\% & 51\% \\ \hline
			Encodji~\cite{C2019} & \multirow{3}{*}{X} & \multirow{3}{*}{X} & \multirow{3}{*}{X} & & 61\% & 43\% & 41\% & 36\%\\
			NN\&Heatmap~\cite{GMS2021FUHL} & & & & & 59\% & 34\% & 38\% & 37\% \\
			ET\&Heatmap~\cite{GMS2021FUHL} & & & & & 60\% & 37\% & 38\% & 38\%\\ \hline
			Encodji~\cite{C2019} & \multirow{3}{*}{X} & & & \multirow{3}{*}{X} & 90\% & 82\% &75\% & 71\%\\
			NN\&Heatmap~\cite{GMS2021FUHL} & & & & & 83\% & 73\% & 70\% & 43\%\\
			ET\&Heatmap~\cite{GMS2021FUHL} & & & & & 88\% & 77\% & 77\% & 51\%\\\hline
			Encodji~\cite{C2019} & & \multirow{3}{*}{X} & & \multirow{3}{*}{X} & 91\% & 81\% &78\% & 70\% \\
			NN\&Heatmap~\cite{GMS2021FUHL} & & & & & 83\% & 74\% & 73\% & 48\% \\
			ET\&Heatmap~\cite{GMS2021FUHL} & & & & & 87\% & 78\% & 76\% & 55\% \\\hline
			Encodji~\cite{C2019} & & & \multirow{3}{*}{X} & \multirow{3}{*}{X} & 91\% & 83\% & 79\% & 74\% \\
			NN\&Heatmap~\cite{GMS2021FUHL} & & & & & 84\% & 76\% & 73\% & 57\% \\
			ET\&Heatmap~\cite{GMS2021FUHL} & & & & & 88\% & 81\% & 77\% & 62\% \\ \hline
			Encodji~\cite{C2019} & \multirow{3}{*}{X} & \multirow{3}{*}{X} & \multirow{3}{*}{X} & \multirow{3}{*}{X} & 92\% & 85\% & 81\% & 76\% \\
			NN\&Heatmap~\cite{GMS2021FUHL} & & & & & 86\% & 79\% & 76\% & 58\% \\
			ET\&Heatmap~\cite{GMS2021FUHL} & & & & & 89\% & 83\% & 78\% & 63\% \\ \hline
			Chance &  & & & & 25.0\% & 12.5\% & 0.48\% & 1.58\% \\ \hline
		\end{tabular}
	\end{table}
	
	\begin{table}[htb]
		\caption{The results of different data generation algorithms for fake scan path generation. Abbreviations: NN is neural network, ET is ensemble of bagged trees, and HOV is histogram of oriented gradients.}
		\label{tbl:evalFake}
		\centering
		\begin{tabular}{llccc}
			\toprule
			& & \multicolumn{3}{c}{Successful deception} \\
			Data & Method & Sim~\cite{fuhl2018simarxiv} & VAE~\cite{FCDGR2020FUHL} & Proposed\\
			\midrule
			\multirow{5}{*}{Challenge~\cite{otero2008saccades,mccamy2014highly}} & Encodji~\cite{C2019} & 32\% & 41\% & 49\% \\
			& NN\&Heatmap~\cite{GMS2021FUHL} & 53\% & 69\% & 75\% \\
			& NN\&HOV~\cite{ICMIW2019FuhlW1} & 57\% & 72\% & 79\% \\
			& ET\&Heatmap~\cite{GMS2021FUHL} & 44\% & 61\% & 71\% \\
			& ET\&HOV~\cite{ICMIW2019FuhlW1} & 44\% & 62\% & 73\% \\ \hline
			\multirow{5}{*}{Hollywood~\cite{costela2019free}} & Encodji~\cite{C2019} & 4\% & 19\% & 31\% \\
			& NN\&Heatmap~\cite{GMS2021FUHL} & 11\% & 37\% & 57\% \\
			& NN\&HOV~\cite{ICMIW2019FuhlW1} & 13\% & 42\% & 62\% \\
			& ET\&Heatmap~\cite{GMS2021FUHL} & 7\% & 33\% & 49\% \\
			& ET\&HOV~\cite{ICMIW2019FuhlW1} & 10\% & 35\% & 55\% \\ \hline
		\end{tabular}
	\end{table}

	Our scan path generator and model creator are implemented in Matlab. The scripts are provided online and everybody can access them via the link in the abstract. For the evaluation we used the implemented machine learning approaches in Matlab as well as we used Matlab to compute the Encodji~\cite{C2019}, heatmap~\cite{GMS2021FUHL}, and histogram of oriented velocities (HOV)~\cite{ICMIW2019FuhlW1} input format. We did not include the rule based learning~\cite{} since it required to many computation time~\cite{ICMIW2019FuhlW2}. The hardware for our evaluation was a desktop PC with an AMD Ryzen 9 3950X 16-Core Processor and 64 GB DDR4 ram. We did not use a GPU, since all models are small enough to train and execute them fast on a CPU. We conducted two experiments. The first is about improving classification accuracy with simulated data, and the second is about generating gaze paths that pretend to be from a person. Alle evaluations are with additional data augmentation. The augmentations used is random scan path cropping, adding random noise to the gaze points, combining random scan path of the same class, and adding up to 5\% of random gaze points. For our evaluation, we performed a 5-fold cross validation. For the ensemble of bagged decision trees (ET), we performed an automatic optimal parameter search from Matlab. For the neural networks (NN), we used a hidden layer with 100 neurons and determined the optimal training parameters via a grid search. For the Encodeji procedure, we used the parameters and the model from the paper \cite{C2019}.
	
	The VAE ~\cite{FCDGR2020FUHL} simulator was trained on the training data for each class and random number vectors were used to generate 1000 simulated gaze data for each class. We used the same procedure for our approach whereby data augmentation was used during training for the VAE as well as our approach. In the case of the \cite{fuhl2018simarxiv} (Sim), gaze paths were generated using the gaze data and the stimulus. For each procedure, 1000 gaze paths were generated for each class and used as additional training data in the first experiment. In the second experiment, the generated gaze data was used to fool the machine learning methods.
	
	\textbf{The ETRA2019Challenge data~\cite{otero2008saccades,mccamy2014highly}} consists of 960 recordings with 45 seconds duration each. The data was collected from 8 subjects. The performed tasks are fixation, visual search and visual exploration. The eye tracking device used is an EyeLink II and during recording the subjects had to use a chin-rest to increase the accuracy of the eye tracker. In total, there were 4 different stimuli used. In our experiment, we use the stimulus and subject classification.
	
	\textbf{The Hollywood data~\cite{costela2019free}} provides eye tracking data of 63 subjects watching Hollywood video clips. Each clip had a length of 30 seconds and in total there are 206 different video clips. Each subject watched $\approx40$ clips. The used Eye Tracker was an Eyelink 1000. In addition to the subject and stimuli information, there is also a rating given for each clip, as well as if the movie was seen before. In this work, we use the stimulus as well as the subject for our classification experiments.
	
	Looking at Table~\ref{tbl:evalCh} it becomes clear that none of the simulators alone comes close to the results of the real data. If all simulators are used together, the individual results improve, but the real data are still significantly better. If the data of the simulators are used together with the real data, the best results are obtained. Our approach provides the largest contribution, but it is not drastically different from the other simulators. The best results are obtained when all generated data and the real data are used together for training. Here there is a significant improvement compared to the results obtained with the real data only.

	In our machine learning deception experiment (table~\ref{tbl:evalFake}), we additionally used histograms of oriented velocities~\cite{ICMIW2019FuhlW1}. These do not use spatial position, only orientations. This makes them perform even worse than the heatmap features in the pure classification experiment. As seen in table~\ref{tbl:evalFake}, it is noticeable that the methods based on the heatmap and HOV features are the easiest to fool. This is due to the fact that these methods quantize the data. In the case of Encodji coding, the deception becomes much more difficult. Overall, our method performs best in this experiment, which is most evident in the Hollywood dataset.


	\section{Limitations}
	The limitations of our algorithm are on the one hand that it is not necessarily the case that real human behavior is simulated. This is due to the fact that our algorithm learns the hierarchy of scan paths from humans and also their distribution, but combines this later randomly. The problem here is that individual parts of the generated scan path are definitely human, but the combination does not necessarily reflect human behavior. Also, nowadays, it is not possible to prove or determine what exactly in a human scan path are the characteristic features of a person and which are not. Another limitation of this work is the evaluation regarding the faking of human scan paths (Table~\ref{tbl:evalFake}). In this evaluation it can be evaluated how many similar scan paths our simulator has generated, which have faked the machine learning approach, but not whether these are good simulated human scan paths regarding the human behavior.

	\section{Conclusion}
	In this paper, we have presented a new fully data-based scan path generation algorithm that can learn from human scan paths. In the first step, the algorithm generates a hierarchy using K-Means clustering and then computes the principal components using the distribution of the next clusters. During generation, the hierarchy is processed, and new clusters are generated in each generated layer based on the principal components. This combines the behavior of from the learned data so that all parts of the generated scan path are human, but the combination does not necessarily result in a human scan path. In our evaluation, we looked at combining different scan path generators as well as real data to train scan path classifiers. Here we could show that the presented generator is at least as good as the state of the art. In combination with the generated data of the other generators and the real data, the best result was obtained, which means that our approach extends the previous possibilities to obtain training data. Furthermore, we conducted experiments to simulate human scan paths to fool a machine learning procedure. Here our method performed best. Overall, we hope that the presented approach is a helpful contribution to the state of the art and helps further researchers to improve scan path classification as well as scan path generation.

	\bibliographystyle{plain}
	\bibliography{template}

\end{document}